\documentclass [prd,nofootinbib]  {revtex4}
\usepackage{amssymb}

\def\no{\, : \,}

\newcommand{\be}{\begin{equation}}
\newcommand{\ee}{\end{equation}}
\newcommand{\bea}{\begin{eqnarray}}
\newcommand{\eea}{\end{eqnarray}}

\begin{document}

\title{%
Star product and interacting fields on $\kappa$-Minkowski space}
\author{Jerzy Kowalski-Glikman}
\email{jurekk@ift.uni.wroc.pl}
\author{Adrian Walkus}
\email{walkus@ift.uni.wroc.pl}
\affiliation{Institute for Theoretical Physics, University of Wroc\l{}aw, Pl.\ Maxa Borna 9, Pl--50-204 Wroc\l{}aw, Poland.}

\begin{abstract}
In this note we extend the methods developed by Freidel et.\ al.\ \cite{Freidel:2006gc} to derive the form of $\phi^4$ interaction term in the case of scalar field theory on $\kappa$-Minkowski space, defined in terms of star product. We present explicit expressions for the  $\kappa$-Minkowski star product. Having obtained the the interaction term we use the resulting deformed conservation rules to investigate if they lead to any threshold anomaly, and we find that in the leading order they do not, as expected.
\end{abstract}

\maketitle

Field theory on the non-commutative $\kappa$-Minkowski space \cite{Majid:1994cy}, \cite{Lukierski:1993wx}, defined by the relation
\begin{equation}\label{1}
 [\hat x_0, \hat x_i] = - \frac{i}{\kappa}\, \hat x_i.
\end{equation}
has recently attracted more and more attention. There are many reasons to be interested in such a theory. Firstly, as shown in \cite{Majid:1994cy} $\kappa$-Minkowski space is a dual of the momentum sector of $\kappa$-Poincar\`e algebra \cite{Lukierski:1991pn}, \cite{Lukierski:1992dt}, which is an interesting example of a Hopf algebra, being a quantum deformation of the standard Poincar\`e algebra. It has been speculated that this deformed algebra may play a role in description of kinematics of ultra-relativistic particles, which would require an additional observer--independent scale of dimension of mass (dubbed Doubly Special Relativity) \cite{Amelino-Camelia:2000ge}, \cite{Amelino-Camelia:2000mn}, \cite{KowalskiGlikman:2001gp}, \cite{Bruno:2001mw}. Further, there are arguments that the very reason for the space-time symmetry algebra deformations and emergence of the non-commutative space are quantum gravity effects (see, eg.\ \cite{Kowalski-Glikman:2006vx}, \cite{Girelli:2009yz}), similar to those established in the context of 2+1 gravity \cite{Freidel:2005bb}, \cite{Freidel:2005me}.

There exists an extensive literature concerning construction of (mostly scalar) classical field theories on $\kappa$-Minkowski space \cite{Kosinski:1999ix}, \cite{Amelino-Camelia:2001fd}, \cite{Kosinski:2003xx}, \cite{Daszkiewicz:2004xy}, \cite{Dimitrijevic:2004nv}, \cite{Freidel:2006gc}, \cite{Freidel:2007hk}, \cite{Freidel:2007yu} as well as their quantization (see eg.\ \cite{Arzano:2007ef}, \cite{Daszkiewicz:2007ru}, \cite{Young:2007ag} and references therein.) However these papers are devoted mostly to the analysis of free fields and surprisingly little is known about their interactions, even in the simplest case of a scalar field with $\phi^4$ interaction (to our knowledge this issue has been addressed only in the papers \cite{Kosinski:1999ix}, \cite{Amelino-Camelia:2001fd}, \cite{Daszkiewicz:2004xy}). In this paper we would like to at least partially fill this gap. To do that we will make use of the star product formalism, using the techniques worked out in \cite{Freidel:2006gc}.

Let us start with recalling some facts from \cite{Freidel:2006gc}. The first observation is that since (\ref{1}) are defining relations of a Lie algebra $an_3$ (called sometimes Borel algebra) we can regard the ordered plane waves on $\kappa$-Minkowski space
\begin{equation}\label{2}
\hat{e}_k \equiv \no e^{i k_\mu \hat{x}^\mu}\no =  e^{i \mathbf{k}\,
\hat \mathbf{x}}\, e^{-ik_0 \hat x_0}.
\end{equation}
as elements of the corresponding Lie group. Therefore the composition of the plane waves,
\begin{equation}\label{3}
    \hat{e}_k \hat{e}_l = \hat{e}_{kl}, \quad  kl\equiv (k_0 + l_0, k_i +
e^{-\frac{k_0}{\kappa}}\, l_i)
\end{equation}
leading to the deformed composition of momenta rule, which is usually understood in terms of the coproduct of $\kappa$-Poincar\`e algebra, can be now interpreted as a simple consequence of the $AN_3$ group structure.

The key point of the construction presented in \cite{Freidel:2006gc} was to define a map, called the Weyl map, that translates an algebra of function (fields) on $\kappa$-Minkowski space with product defined by (\ref{3}), to the algebra of functions on the ordinary Minkowski space ${\cal M}^4$, equipped with star product  replacing the standard multiplication. Under this map the ordered plane wave (\ref{2}), labeled by momentum $k$ transforms into an ordinary plane wave labeled by momentum $P$ (whose components form a Lorentz vector), to wit
\begin{equation}\label{4}
  e^{i \mathbf{k}\,
\hat \mathbf{x}}\, e^{-ik_0 \hat x_0} \mapsto e^{iP(k)_\mu\, x^\mu}, \quad x^\mu\in {\cal M}^4
\end{equation}
where
\begin{eqnarray}
 {P_0}(k_0, \mathbf{k}) &=&  \kappa\sinh
{\frac{k_0}{\kappa}} + \frac{\mathbf{k}^2}{2\kappa}\,
e^{  \frac{k_0}{\kappa}} \nonumber\\
 P_i(k_0, \mathbf{k}) &=&   k_i \, e^{  \frac{k_0}{\kappa}}
 \label{5}
\end{eqnarray}
It is worth reminding at this point that $P_\mu$ above along with
\begin{equation}\label{6}
    {P_4}(k_0, \mathbf{k}) =  \kappa\cosh
{\frac{k_0}{\kappa}} - \frac{\mathbf{k}^2}{2\kappa}\,
e^{  \frac{k_0}{\kappa}}
\end{equation}
are coordinates on the elliptic de Sitter momentum space $P_{4}^{2}=
\kappa^2 -\mathbf{P}^{2} +P_{0}^{2} $ and that on-shell $-\mathbf{P}^{2} +P_{0}^{2}=m^2$ we have $P_{4}=\sqrt{\kappa^2+m^2}$.

The star product counterpart of (\ref{3}) reads
\begin{equation}\label{7}
    e^{iP(k)\cdot x} \star e^{iP(l)\cdot x} \equiv e^{iP(kl)\cdot x}
\end{equation}
and can be related to the coproducts of momenta $P$ of $\kappa$-Poincar\`e algebra, which can be found \cite{Freidel:2007hk}. To see this explicitly let us rewrite (\ref{7}) in the form
\begin{equation}\label{7.1}
    e^{iP\cdot x} \star e^{iQ\cdot x} \equiv e^{i(P\oplus Q)\cdot x}
\end{equation}
where by direct calculation one finds
\begin{equation}\label{7.2}
    (P\oplus Q)_0 =  \frac1\kappa\, P_{0}Q_{+}+ \kappa\, \frac{Q_{0}}{P_{+}} + \frac{\mathbf{P}\cdot \mathbf{Q}}{P_{+}} ,  \quad
(P\oplus Q)_i = \frac1\kappa\, P_{i} Q_{+} + Q_{i}
\end{equation}
with $$P_{+} =P_{4} + P_{0} \equiv \sqrt{\kappa^2+P_0^2-\mathbf{P}^2}+ P_0$$ and analogous expression for $Q_+$.  Notice that in the leading order in $1/\kappa$
$$ (P\oplus Q)_0 = P_0+Q_0 +O\left(\frac1\kappa\right), \quad (P\oplus Q)_i = P_i+Q_i +O\left(\frac1\kappa\right)$$
as it should be.

Using formulas (\ref{7.1}) and (\ref{7.2}), by direct calculation, one can derive an explicit formula for the star product of two functions, to wit
\begin{equation}\label{7.3}
\psi(x)\star \phi(x)=\lim_{a,b \to 0 }\psi(x+a)\exp\left[-ix_0\Delta_0\left(\overleftarrow{ \frac{\partial}{\partial a}}\,\stackrel{\longrightarrow}{ \frac{\partial}{\partial b}} \right) \right]\exp\left[i\mathbf{x\,\Delta}\left(\stackrel{\longleftarrow}{ \frac{\partial}{\partial a}}\,\overrightarrow{ \frac{\partial}{\partial b}} \right) \right]\phi(x+b)
\end{equation}
where
\begin{equation}\label{7.4}
\Delta_0\left(\stackrel{\longleftarrow}{ \frac{\partial}{\partial a}}\,\overrightarrow{ \frac{\partial}{\partial b}} \right)=\frac{1}{i}\stackrel{\longleftarrow}{ \frac{\partial}{\partial a^0}}\, \overrightarrow{ E}+\stackrel{\longleftarrow}{E^{-1}}\frac{1}{i}\stackrel{ \longrightarrow}{ \frac{\partial}{\partial b^0}}+\frac{\stackrel{\longleftarrow}{E^{-1}}}{\kappa} \left(\frac{1}{i}\stackrel{\longleftarrow}{ \frac{\partial}{\partial \mathbf{a }}} \right)\left(\frac{1}{i}\overrightarrow{ \frac{\partial}{\partial \mathbf{b }}} \right)- \frac{1}{i}\stackrel{\longleftarrow}{ \frac{\partial}{\partial a^0}}-\frac{1}{i}\overrightarrow{ \frac{\partial}{\partial b^0}}
\end{equation}
and
\begin{equation}\label{7.5}
\mathbf{\Delta}\left(\stackrel{\longleftarrow}{ \frac{\partial}{\partial a}}\,\overrightarrow{ \frac{\partial}{\partial b}} \right)=\frac{1}{i}\stackrel{\longleftarrow}{ \frac{\partial}{\partial \mathbf{a}}}\, \overrightarrow{ E}-\frac{1}{i}\overleftarrow{ \frac{\partial}{\partial \mathbf{a}}}.
\end{equation}
In these formulas we introduced the operator
\begin{equation}\label{7.6}
\overleftarrow{ E}=\frac{1}{i\kappa }\overleftarrow{ \frac{\partial}{\partial a^0}}+\sqrt{-\frac{\overleftarrow{ \Box}}{ \kappa^2}+1}, \qquad \overleftarrow{ \Box}=\overleftarrow{ \frac{\partial^2}{\partial a_0^2}}-\overleftarrow{ \frac{\partial^2}{\partial \mathbf{a}^2}}
\end{equation}

In order to formulate field theory we need one more definition, namely that of conjugation
\begin{equation}\label{8}
\left(e^{iP(k)\cdot x}\right)^\dagger\equiv e^{iP(k^{-1})\cdot
x}, \quad k^{-1}=(-k_0, -k_i \, e^{k_0})
\end{equation}
The expression for $P(k^{-1})$ is provided by an antipode, the explicit expressions for whose can be again found in \cite{Freidel:2007hk}. Writing
\begin{equation}\label{8.1}
\left(e^{iP\cdot x}\right)^\dagger\equiv e^{i(\ominus P)\cdot
x}
\end{equation}
we have
\begin{equation}\label{8.2}
    (\ominus P)_0=\frac{\kappa^2}{P_0+P_4}-P_4= -P_{0} +\frac1\kappa\, \frac{\mathbf{P}^{2}}{P_{0}+P_{4}}, \quad (\ominus P)_i= -\kappa\, \frac{  P_i}{P_0+P_4}
\end{equation}
and in the leading order it reproduces the standard minus
$$
(\ominus P)_0=-P_0 +O\left(\frac1\kappa\right), \quad (\ominus P)_i=-P_i +O\left(\frac1\kappa\right)
$$
Below we will need an expression for $(\ominus P)\oplus Q$, which in components reads
\begin{equation}\label{8.2a}
   \left[(\ominus P)\oplus Q\right]_0=-\frac1\kappa\, P_0Q_+ +\frac1\kappa\, Q_0P_++\frac1{\kappa}\, \frac{P_i}{P_+}\left(P_i\, Q_+-Q_i\, P_+\right)
\end{equation}
\begin{equation}\label{8.2b}
   \left[(\ominus P)\oplus Q\right]_i=\frac{P_+Q_i-P_iQ_+}{P_+}
\end{equation}
and
\begin{equation}\label{8.2c}
   \left[(\ominus P)\oplus Q\right]_+=\frac{Q_+}{P_+}
\end{equation}

By direct calculation one can check that the conjugation of the field $\phi$ is given by
\begin{equation}\label{8.3}
\phi^\dag(x)=\lim_{c \to 0}\,\exp\left\{-ix_0\left[S_0\left(\frac{\partial}{\partial c} \right)-\frac{1}{i}\frac{\partial}{\partial c^0} \right] +i\mathbf{x}\left[\mathbf{S}\left(\frac{\partial}{\partial c} \right)-\frac{1}{i}\frac{\partial}{\partial \mathbf{c}} \right] \right\}\phi^*(x+c)
\end{equation}
where $\psi^*$ is the standard complex conjugation of $\psi$ and
\begin{equation}\label{8.4}
S_0\left(\frac{\partial}{\partial c} \right)=\kappa^2\left(-\frac{1}{i}\,\frac{\partial}{\partial c^0}+\sqrt{-\Box+\kappa^2} \right)^{-1}-\sqrt{-\Box+\kappa^2}
\end{equation}
and
\begin{equation}\label{8.5}
\mathbf{S}\left(\frac{\partial}{\partial c} \right)=\frac{\kappa}{i}\frac{\partial}{\partial \mathbf{c}}\left(-\frac{1}{i}\,\frac{\partial}{\partial c^0}+\sqrt{-\Box+\kappa^2} \right)^{-1}.
\end{equation}
Using both expressions, for star product and conjugation, one can easily derive the formula for the product of the form $\psi^\dagger \star \phi$. This formula is rather complex, so it comes as a pleasant surprise that it has the form
\begin{equation}\label{8.6}
    \psi^\dagger \star \phi = \psi^* \sqrt{1-\Box/\kappa^2} \phi + \mbox{total derivative}
\end{equation}
This is the reason why the action for the free scalar field in star product formalism, presented in \cite{Freidel:2006gc}, has such a nice, compact form. Unfortunately, as we will see below this simplicity of free theory does not carry over into the interacting one.\newline

The conjugate plane wave can be used to perform Fourier transform; explicitly for the field $\phi(x)$ its transform reads
\begin{equation}\label{9}
    \tilde{\phi}(P(k))= \int \mathrm{d}^4x
\left(e^{iP(k)\cdot x}\right)^\dagger \star {\phi}(x)
\end{equation}
One checks that this transform is related to the standard one by
\begin{equation}\label{10}
    \tilde{\phi}(P)= \frac{P_4}{\kappa}\,
\int \mathrm{d}^4x \, e^{-iP\cdot x}  {\phi}(x), \quad \phi(x)= \int
\frac{\mathrm{d}^4P}{(2\pi)^4} \frac{e^{iP x}}{P_4/\kappa}
\tilde{\phi}(P)=\int
d\mu(P) {e^{iP x}}
\tilde{\phi}(P)
\end{equation}

Equipped with these technical tools we can now proceed to construction of the action for an interacting scalar field. Consider first the quadratic term, which was evaluated in \cite{Freidel:2006gc}. It reads
$$ \int d^4x   \phi^\dagger \star \phi =\int d^4x \int d\mu(P) d\mu(Q) \tilde\phi^*(P)\, \tilde\phi(Q) \left(e^{iPx}\right)^\dagger \star e^{iQx}$$\begin{equation}\label{11} =\int d^4x \int d\mu(P) \int d\mu(Q) \tilde\phi^*(P)\, \tilde\phi(Q)\, e^{i(\ominus P\oplus Q)x}
\end{equation}
It turns out that the space-time integral of the exponential factor in this formula can be greatly simplified being essentially proportional to the standard momentum delta function, as follows.  Using (\ref{8.2a}) and (\ref{8.2b}) one easily checks
$$
    e^{i(\ominus P\oplus Q)x}=\exp\left[ix^0\left(-\frac1\kappa\, P_0Q_+ +\frac1\kappa\, Q_0P_++\frac1{\kappa}\, \frac{P_i}{P_+}\left(P_i\, Q_+-Q_i\, P_+\right)\right)\right]\,\exp\left[ix^i\left(\frac{P_+Q_i-P_iQ_+}{P_+}\right)\right]
$$
which by changing variables in integral $x^i \rightarrow x^i +x^0\, P^i/\kappa$ can be simplified to
\begin{equation}\label{12}
 e^{i(\ominus P\oplus Q)x}=\exp\left[ix^0\left(-\frac1\kappa\, P_0Q_+ +\frac1\kappa\, Q_0P_+\right)\right]\,\exp\left[ix^i\left(\frac{P_+Q_i-P_iQ_+}{P_+}\right)\right]
\end{equation}
In the next step we use the relation $2P_0 =P_+-\kappa^2 P_+^{-1} +\mathbf{P}^2/P_+$ to obtain
$$
-\frac1\kappa\, P_0Q_+ +\frac1\kappa\, Q_0P_+=-\kappa\, \frac{P_+^2-Q_+^2}{2P_+Q_+} +\frac1\kappa\, \frac{(P_+\mathbf{Q})^2-(Q_+\mathbf{P})^2}{2P_+Q_+}
$$
where the last term can be again removed by change of variables similar to that above. Then one rescales $x_0$ and changes momentum variables in exponent to obtain
\begin{equation}\label{12a}
 e^{i(\ominus P\oplus Q)x}\sim\frac{2P_+Q_+}{\kappa(P_++Q_+)} \, \exp\left[ix^0\left(- P_+ + Q_+\right)\right]\,\exp\left[ix^i\left({Q_i-P_i\frac{Q_+}{P_+}}\right)\right]
\end{equation}
where the $\sim$ indicates that the equality holds only in the integral (\ref{11})

Let us now generalize (\ref{11}) to the case of an interacting field with the $\phi^4$ interactions. Then we have
$$
\int d^4x   \left(\phi^\dagger \star \phi \right)^\dagger \star \left(\phi^\dagger \star \phi \right) $$
\begin{equation}\label{13}
= \int d^4x  d\mu(P)  d\mu(Q)  d\mu(R) d\mu(S) \tilde\phi(P)\, \tilde\phi^*(Q) \tilde\phi^*(R)\, \tilde\phi(S)\left(\left(e^{iPx}\right)^\dagger \star e^{iQx}\right)^\dagger \star \left(e^{iRx}\right)^\dagger \star e^{iSx}
\end{equation}
Using the expressions (\ref{12a}) and (\ref{8.2b}), (\ref{8.2c}) we express the star product of exponents as follows
$$
\left(\left(e^{iPx}\right)^\dagger \star e^{iQx}\right)^\dagger \star \left(e^{iRx}\right)^\dagger \star e^{iSx}
$$
\begin{equation}\label{14}  \sim2\frac{Q_+S_+}{Q_+R_++S_+P_+}\,\exp\left[ix^0\kappa\left(-\frac{Q_+}{P_+}+\frac{S_+}{R_+}\right)\right]
\exp\left[ix^i\left(S_i-R_i\,\frac{S_+}{R_+}-Q_i\frac{S_+}{R_+}\frac{P_+}{Q_+}+P_i\,\frac{S_+}{R_+}\right)\right]
\end{equation}
This formula along with (\ref{13}) defines the quartic interaction term. Integrating over space-time we can simplify a bit, and the final expression reads
$$
\int d^4x   \left(\phi^\dagger \star \phi \right)^\dagger \star \left(\phi^\dagger \star \phi \right) $$
\begin{equation}\label{15}
    =\int  d\mu(P)  d\mu(Q)  d\mu(R) d\mu(S) \tilde\phi(P)\, \tilde\phi^*(Q) \tilde\phi^*(R)\, \tilde\phi(S) \left(\frac{S_+}{R_+}\right)\, \delta\left(-\frac{Q_+}{P_+}+\frac{S_+}{R_+}\right)\, \delta\left(S_i-R_i\,\frac{S_+}{R_+}-Q_i+P_i\,\frac{S_+}{R_+}\right)
\end{equation}
As observed in the similar context in \cite{Amelino-Camelia:2001fd} and \cite{Daszkiewicz:2004xy} the symmetry between $Q$ and $R$, and $P$ and $S$ enforces symmetrization of delta functions. Thus the integral (\ref{15}) describes four different processes (four different conservation rules), which seems to be a generic feature of interactions in the theories with non-trivial coproduct. Therefore, although the conservation rule associated with the vertex is not symmetric under exchange of momenta, there is not one but four conservations, each equally contributing to quantum probability of the scattering process, which together render the amplitude symmetric with respect to the particles exchange.

This formula makes it possible to formulate deformed energy and momentum conservation rules for our theory. The energy conservation reads
\begin{equation}\label{16}
    Q_+R_+=P_+S_+
\end{equation}
and for particles on shell
\begin{equation}\label{17}
    Q_+=Q_0+\kappa\sqrt{1+\frac{m_Q^2}{\kappa^2}}
\end{equation}
with similar expressions for other particles. Notice that this formula is symmetric under exchange of particles imposed by the form of (\ref{15}).

In (\ref{17}) we make an obvious generalization of the above result allowing four momenta $P,Q,\ldots$ to label particles of different masses. It should be noted that since masses of the particles are by many orders of magnitude smaller than the scale $\kappa$ for all practical purposes we can rewrite (\ref{17}) as
\begin{equation}\label{18}
    Q_+=Q_0+\kappa
\end{equation}
In this case (\ref{16}) becomes
\begin{equation}\label{19}
    Q_0+R_0+\frac{Q_0R_0}\kappa=P_0+S_0+\frac{P_0S_0}\kappa
\end{equation}
Let us now turn to the momentum conservation
\begin{equation}\label{20}
    Q_i+R_i\,\frac{Q_+}{P_+}=S_i+P_i\,\frac{S_+}{R_+}
\end{equation}
Unlike (\ref{16}) this formula is not symmetric so it should be accompanied by three more conservation laws, corresponding to different channels
\begin{eqnarray}
% \nonumber to remove numbering (before each equation)
 R_i+Q_i\,\frac{R_+}{P_+} &=& S_i+P_i\,\frac{S_+}{Q_+} \nonumber\\
  Q_i+R_i\,\frac{Q_+}{S_+} &=& P_i+S_i\,\frac{P_+}{R_+} \label{21}\\
  R_i+Q_i\,\frac{R_+}{S_+} &=& P_i+S_i\,\frac{P_+}{Q_+}\nonumber
\end{eqnarray}
To the leading order in $1/\kappa$ (\ref{20}) takes the form
\begin{equation}\label{22}
    Q_i+R_i\left(1+\frac{Q_0}\kappa -\frac{P_0}\kappa\right)=S_i+P_i\left(1+\frac{S_0}\kappa -\frac{R_0}\kappa\right)
\end{equation}
To see these conservation rules in action let us consider the process of electron-positron pair creation $\gamma+\gamma \rightarrow e^++e^-$ with one very hard and one soft photon. Let us denote the energies of hard and soft photons by ${\cal E}$ and $\epsilon$, respectively. Since we will be interested in threshold condition for head-to-head scattering of photons only, we can assume that energies and momenta of the electron and the positron are the same, and equal $e$ and $p$ respectively. Using (\ref{19}) and (\ref{22}) and taking into account that the hard photons threshold energy ${\cal E}_{th} \gg \epsilon$ we find
\begin{equation}\label{23}
    {\cal E}_{th} + \epsilon =2e+\frac{e^2}\kappa
\end{equation}
\begin{equation}\label{24}
    {\cal E}_{th} - \epsilon =2p+\frac{pe}\kappa
\end{equation}
One can check by inspection that all other forms of the momentum conservation rule (\ref{21}) lead, in the leading order, to the same resulting equation. Using the dispersion relation for electron/positron, $e^2=p^2+m^2$ with $m$ being the electron mass we easily find
\begin{equation}\label{25}
   {\cal E}_{th}\epsilon=m^2
\end{equation}
Not surprisingly we find therefore that the deformed  conservation rules lead, in the leading order in $1/\kappa$ to the standard, undeformed threshold condition. This is, as argued in  \cite{AmelinoCamelia:2003ex}, a rather generic feature of DSR theories, where the deformation is by construction covariant with respect to the (deformed) action of the Lorentz group. Our theory will, of course, lead to deformations, but they, in agreement with \cite{AmelinoCamelia:2003ex}, will be very small.

\section*{Acknowledgment} For JKG  this research was supported in part  by research projects N202 081 32/1844 and NN202318534.

\end{document}